\documentclass{eas}
\usepackage{graphicx}
\newcommand{\mo}    {M_{\odot}}

\newcommand{\pri}   {${\rlap.}^{\prime \prime}$}
\newcommand{\rl}    {${\rlap.}^{\rm s}$}

\begin{document}

\title{Stellar radio astrophysics} 
\author{J.~M.~Paredes}\address{Departament d'Astronomia i Meteorologia, Universitat de
Barcelona, Av. Diagonal 647, 08028 Barcelona, Spain (jmparedes@ub.edu)}

\begin{abstract}
Radio emission has been detected from all the stages of stellar evolution
across the HR Diagram. Its presence reveals both astrophysical phenomena and
stellar activity which, otherwise, would not be detectable by other means. The
development of large, sensitive interferometers has allowed us to resolve the
radio structure of several stellar systems, providing insights into the mass
transfer process in close binary systems. I review the main characteristics of
the radio emission from several kinds of stars, paying special attention to
those cases where such an emission originates in relativistic jets. 
\end{abstract}
\maketitle
\section{Introduction}
If the stellar radio emission was restricted to solar-like activity levels, 
our knowledge of the Galaxy in radio wavelenghts would be significantly
lesser. The radio flux density $S_{\nu}$ at a frequency $\nu$ of a star,
measured at the earth, can be expressed in terms of the Rayleigh-Jeans
approximation to the black body formula 
\begin{equation}
S_{\nu} = {2 k \nu^{2} T_{\rm B} \Omega_{\rm S} \over c^{2}},
\end{equation}
where $k$ is the Boltzmann's constant, $T_{\rm B}$ is the brightness
temperature and $\Omega_{\rm S}$ is the solid angle subtended by the star. The
maximum distance at which a star like the Sun would be observable at 5~GHz
with a radio telescope, assuming a typical minimum flux density of 1~mJy, is
0.2~pc for the quiet Sun, 0.3~pc for the slowly varying component, and 7.6 pc
for the very strong solar radio bursts. If we take into account that the
nearest star, Proxima Centauri, is 1.3~pc away, the solar-like emission would
be observed only from a few nearby stars. In fact, the first {\it radio stars}
detected were not really stars emitting at radio wavelengths but strong
extragalactic sources.

Fortunately, the situation is much better. Since $S_{\nu} \propto T_{\rm B}
\Omega_{\rm S}$, it is possible to detect radio stars that have either
unusually large emitting surfaces, generally associated to stars having
massive stellar winds, or enhanced brightness temperatures, associated to
stars exhibiting  powerful non-thermal radio emission. 

Early efforts to observe radio stars concentrated on the nearby red
supergiants. Kellermann \& Pauliny-Toth (\cite{kellerman66}) reported a signal
of 0.11$\pm$0.03~Jy while observing Betelgeuse at 2.8~cm, although no signal
was detected on the next 11 nights. Seaquist (\cite{seaquist67}) reported the
detection of Betelgeuse, with a flux density of 0.023$\pm$0.006~Jy at 2.8~cm,
and $\pi$Aurigae, with a flux density of 0.031$\pm$0.009~Jy. However, none of
these stars has been detected at such flux levels since then. The results
obtained in these pioneering observations could have been affected by
different factors, such as unknown problems with the equipment, the intrinsic
variability of most radio stars, and confusion with background sources.

The fact that most of the solar radio emission occurs in a large variety of
flares of varying lifetimes inspired the study of radio emission from
non-solar type stars. Red dwarf stars were known to exhibit flares at optical
wavelengths, and became obvious candidates for radio flare searches. In 1969
Bernard Lovell, at Jodrell Bank, pioneered these observations. The first
indisputable observations of radio flare star emission were the detections of
a radio flare in YZ~CMi and in UV~Ceti at 408~MHz, obtained using long
baseline interferometers (Davis \etal\ \cite{davis78}). The construction of
large aperture synthesis arrays in the mid 1970s allowed to perform reliable
measurements of radio emission and marked the beginning of the stellar radio
astronomy.

The advent of radio interferometry prompted a significant advance in the
knowledge of radio stars. Thanks to this technique it was possible to obtain
a) accurate positions, which allow to see if the position of a radio source
coincides with that of a star, b) higher sensitivity, because all the data
collected for long observing times can be added together, and c) a better
determination of the variability, since a time-variable event is most certain
if independent subsets of an array are detecting the same time-variable event.
Since then, the field of stellar radio astronomy has advanced explosively. We
are now able to image radio emission from stellar sources with unprecedented
resolution and sensitivity. Thanks to these observational advances, the number
and classes of stars known to be sources of radio emission have increased
dramatically. As we will see, radio emission has been detected from all the
stages of stellar evolution, from birth to death, and has revealed
astrophysical phenomena and stellar activity that are not detectable by other
means.

To illustrate the progress made in the field of stellar radio astrophysics, it
is worth noting that two dedicated conferences were conducted in the past two
decades. In 1984, a workshop on stellar continuum radio astronomy was held in
Boulder (Hjellming \& Gibson \cite{hjellming85}). That workshop reviewed the
observational and theoretical work on radio stars from the first decade of the
interferometer era. In 1995, a workshop was held in Barcelona, providing an
overview of the rich and powerful sets of new radio observations accumulated
over the past decade and new theoretical ideas and models arising from them
(Taylor \& Paredes \cite{taylor96}).

\section{Types of radio stars}

\begin{figure} 
\center
\includegraphics[width=7cm]{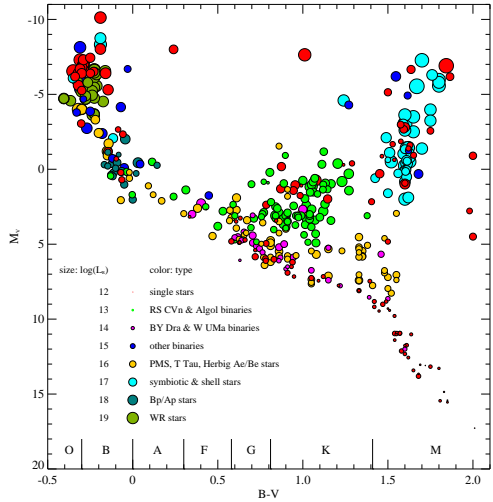}
\caption{HR diagram showing the distribution of several types of stars (from 
G\"udel \cite{gudel02}).} 
\label{hrd}
\end{figure}

The catalogue of Wendker (\cite{wendker95}) of radio continuum emission from
stars contains 3021 systems: 821 were detected at least once, and 2192 have
only upper limits. A first look at this catalogue evidences both the
variability and the low level of radio emission from the stars. The
Hertzprung-Russell diagram for a subset of stellar detections is shown in
Figure~\ref{hrd}. The radio luminosity is indicated by the size of each
circle. Radio emission has been detected from all the stages of stellar
evolution across the HR diagram. The evolutionary state of a star in the
diagram is closely related to its radio activity. Most of the main sequence
and sub-giant objects are non-thermal emitters. In contrast, most of the
giants and many O-B stars are thermal emitters, but they can be detected
because of their large sizes. Novae and X-ray binaries are not shown since the
active source is related to accretion onto a white dwarf and a neutron star or
black hole, respectively, and not directly to their evolutionary state.

The radio stars can be classified according to the underlaying cause of their
enhanced radio luminosity (see also Seaquist \cite{seaquist93}). We can
distinguish the following classes:

\subsection{Stars undergoing mass loss}

{\it OB and Wolf-Rayet stars}: These stars produce free-free (FF) emission
associated to an optically thick wind. Non-thermal emission is also observed,
due either to relativistic electrons embedded in the wind produced by Fermi
acceleration in shock fronts or to collision between the winds of the two
components of a binary.

{\it Be stars}: The FF emission is produced as a consequence of mass loss
through an equatorial disk by centrifugal effects caused by the rapid rotation
of the star.

{\it Single red giants and supergiants}: The winds of these evolved stars are
cool and only partially ionized. They are detectable if the star is relatively
nearby.

{\it VV Cephei stars}: Binaries containing a cool supergiant and a main
sequence B companion. The FF emission comes from a subregion of the supergiant
wind ionized by UV photons from the hotter companion. 

{\it Pre-main sequence stars}: T Tauri stars that can be subdivided into two
classes: classical T Tauri stars (CTT), which emit FF emission from ionized
gas, and weak-lined T Tauri stars (WTT), which emit non-thermal radio emission
from magnetically active regions.

\subsection{Stars exhibiting enhanced solar activity}
 
The sun exhibits non-thermal emission and flare activity associated with
high-energy particles in the chromosphere and corona. The same phenomenon on
larger scales is characteristic of many cool (primarily K-M) stars on and
above the main sequence. The energy supply is thought to be the release of
stellar magnetic field energy by reconnection of field lines. The magnetic
fields may be generated by a dynamo mechanism.

{\it Flare stars}: Single stars that exhibit intense flares from X-rays to
radio waves. The coronal gas is bound by magnetic loops of several 1000~G
covering most of the stellar surface. The stellar flares are due to coherent
emission (at lower frequencies) and incoherent gyrosynchrotron emission (at
higher frequencies). Prototype stars are UV~Ceti, YZ~CMi and AD~Leo.

{\it Close binaries}: The typical components of the RS Canis Venaticorum
(RS~CVn) binary systems are a solar-type star and a more evolved, cool
subgiant. Their radio emission is generally accounted for by a gyrosynchrotron
emission mechanism. The magnetic activity is enhanced by the high rotation
rate of the active subgiant, which is synchronised to the period of orbital
revolution by tidal coupling. Particle acceleration may also occur in the
intrabinary plasma if the interaction between the fields of the two stars
produce reconnection. These mechanisms may also occur in semi-detached systems
(Algol binaries) and in contact binaries (W Ursae Majoris stars). The radio
properties of Algols are similar to those of RS CVn systems, while W UMa stars
are less luminous. The characteristics of the RS~CVn systems are discussed in
more detail in Section~\ref{rscvn}.

{\it Pre-main sequence stars}: Some of the properties of WTT stars are similar
to those of RS CVn stars. The mechanism for producing the radio emission is 
gyrosynchrotron emission from starspot regions.

\subsection{Chemically peculiar stars}

CP stars or Bp-Ap stars: Main sequence stars characterized by over and under
abundances of certain chemical elements and by strong (1-10 kG) dipolar
magnetic fields. There are no convection motions in their stellar envelopes,
and the magnetic field is thought to be a fossil remnant of the dynamo fields
generated in the pre main-sequence phase. Their radio properties (flat radio
spectrum, circular polarization, non-thermal $T_{\rm B}$) are consistent with
gyrosynchrotron emission.

\subsection{Radio emission related to mass transfer in binaries}

The transfer process may involve Roche lobe overflow or stellar wind accretion
from a normal star to a white dwarf (WD), neutron star (NS) or black hole
(BH).

{\it Cataclysmic variables}: Classical novae are semi-detached binaries
containing a main-sequence star and a WD. Nova outbursts (intervals of
$\sim$10$^{4}$~yr) occur when the accreted H-rich material accumulates on the
outer surface of the WD and leads to a thermal runaway and explosive ejection
of the outer envelope. FF emission comes from the expanding ionized ejecta. It
is observed synchrotron radiation from particles accelerated in a shock
(within the ejecta or in the interaction between the nova ejecta and a dense
gas cloud). The magnetic cataclysmic variables (DQ Her and AM Her systems)
contain a late-type MS and a magnetic WD, and the mass transfer process is
modified by the presence of strong magnetic fields of 10$^{5}$--10$^{7}$~G. It
has been observed highly variable flare-like non-thermal radio emission.
Symbiotic stars are interacting binaries containing a cool (red) giant and a
hot companion. These systems emit FF emission from a circumstellar ionized
envelope.

{\it X-ray binaries}: The X-ray emission is produced by accretion of matter
onto a compact companion. See more details in Section~\ref{xrb}.

\section{Linking the radio and optical reference frames}

The determination of a suitable celestial reference frame is of fundamental
importance in astronomy, in particular when dealing with the highly increasing
accuracy of the observations. The International Celestial Reference System
(ICRS) is a quasi-inertial reference system based upon the positions of a
number of extragalactic objects. The International Celestial Reference Frame
(ICRF) is the radio realization of the ICRS, and it is defined by the
positions of 212 extragalactic radio sources derived from VLBI observations
(Ma \etal\ \cite{ma98}). The ICRF is the most precise reference system ever
materialized. On the other hand, the Hipparcos astrometry space mission has
observed $\sim$120\,000 objects evenly distributed over the sky, providing
astrometric parameters with a precision of $\sim$1~mas and
$\sim$1~mas~yr$^{-1}$. The positions and proper motions of the Hipparcos
Catalogue define an optical reference frame that represents a materialization
of the ICRS at optical wavelenghts.

Thus, to determine the best celestial reference frame for astronomical
purposes, it becomes necessary to link the Hipparcos Reference Frame to the
ICRF through objects whose positions and/or proper motions can be referred to
both systems. The observation of radio stars is one of the methods used for
this link. For these observations, it is important to select targets with
small angular diameters (non-thermal emitters) and bright enough in the radio
band to provide accurate positions. Multiple-epoch phase referenced VLBI
observations of selected radio stars (active non-thermal radio emitters which
are also optically bright) have been conducted as part of an on-going
astrometric program to link the Hipparcos reference frame to the 
extragalactic reference frame (Lestrade \etal\ \cite{lestrade99}). The
observational campaigns performed for this link prompted a significant advance
in the study of radio stars. The preferred radio sources for these
observations were RS~CVn binaries and X-ray binaries. These systems will be
discussed in detail in the following sections.

\section{RS~CVn binaries}\label{rscvn}

RS~CVn are one of the most common kind of radio detected stars. RS~CVn systems
are close, chromospherically active binary systems whose enhanced emission can
be detected over a wide range of the spectral domain, from the X-ray to the
radio region. Most of the phenomena observed in these systems at different
wavelengths are attributed to the presence of magnetic fields generated by a
dynamo mechanism.

In general, the radio emission from RS~CVn binary systems is quite variable,
with luminosity levels in the range 10$^{14}$$-$10$^{19}$
erg~s$^{-1}$~Hz$^{-1}$ at centimeter wavelengths (Morris \& Mutel
\cite{morris88}; Drake \etal\ \cite{drake92}). Typical features of the radio
emission from these systems are: a) low-level (quiescent) emission with
moderately circularly polarized emission (degree of circular polarization
$\pi_{\rm c} \leq$30--40\%) and flat or negative spectrum (spectral index
$\alpha \leq 0$, where $S_{\nu} \propto \nu^{\alpha}$); or b) high-level
(flaring) emission, unpolarized or weakly polarized ($\pi_{\rm c} \leq 10$\%),
and with positive spectral index $\alpha$; or c) high-intensity,
short-duration outbursts with high degrees of circular polarization (see Gunn
\cite{gunn96} for a review on RS~CVn binary systems). The mechanisms generally
invoked to account for the observed radio emission are gyrosynchrotron,
synchrotron and coherent (electron-cyclotron maser or plasma radiation)
processes. For reviews on these radio emission mechanisms see, for instance,
Dulk (\cite{dulk85}) and G\"{u}del (\cite{gudel02}).

Linear polarization is not expected to be present in the radio emission from 
RS~CVn systems due to the large Faraday rotation in the stellar coronae.
Circular polarization measurements are, thus, the main input for the 
understanding of the magnetic fields of these stars. Mutel \etal\
(\cite{mutel87}) studied a small sample of late-type binaries, mainly of the
RS~CVn class, and found that non-eclipsing (low inclination angle) systems
have an average circular polarization in their quiescent emission
significantly larger than that for eclipsing systems. Multifrequency
polarization observations (e.g., Mutel \etal\ \cite{mutel87}; Umana \etal\
\cite{umana93}; White \& Franciosini \cite{white95}; Garc\'{\i}a-S\'anchez
\etal\ \cite{garcia03}) revealed properties of both low-level and high-level
emission sources, showing a reversal in the sense of polarization between
1.4~GHz and 5~GHz for non-eclipsing systems. White \& Franciosini
(\cite{white95}) proposed that weak, highly polarized, coherent plasma
emission may be associated with the polarization inversion observed at low
frequencies. In addition, White \& Franciosini (\cite{white95}) found an
increase of the degree of polarization with increasing frequency at high
frequencies, independently of the shape of the spectrum, contrary to what is
expected according to gyrosynchrotron models.

The large coronal sizes of active RS~CVn systems could be accounted for by the
gradual release of magnetic free energy built up by the interaction of the
stellar magnetic fields of the two components of the binary (Uchida \& Sakurai
\cite{uchida83}). Differential rotation in the stars may produce a twisting of
magnetic flux tubes, and the reconnection of loops can give rise to a extended
corona between the two components. VLBA and VLA observations of the quiescent
emission from UX~Ari by Beasley \& G\"udel (\cite{beasley00}) resolved the
source on linear scales comparable to the projected separation of the
components, which indicates the existence of large magnetic structures. They
also detected circular polarization with a polarization gradient across the
radio source, which suggests the existence of interacting magnetic fields
between the components or a large magnetic loop anchored to one component.

The non-thermal nature of the electron distribution responsible for the
flaring emission from RS~CVn systems seems to be well established. The
quiescent emission has been interpreted in two different ways, based on
different assumptions for the distribution of the population of electrons
responsible for the emission observed, namely gyrosynchrotron emission from a
Maxwellian (thermal) distribution or from a power-law (non-thermal) 
distribution (Drake \etal\ \cite{drake92}; Chiuderi-Drago \& Franciosini
\cite{chiuderi93}). However, a thermal gyrosynchrotron emission model with
uniform magnetic field predicts a spectral index $\alpha = -8$ after the peak,
whereas the spectra observed are much flatter. Thermal emission in a magnetic
field $B$ decreasing as $B \propto r^{-1}$, where $r$ is the distance from the
active star, may reproduce the observed quiescent spectrum, but the magnetic
field structure seems unrealistic. Emission from a non-thermal distribution
appears to be the most plausible explanation for the observed properties of
the quiescent emission.

\begin{figure}
\center 
\includegraphics[width=5.5cm]{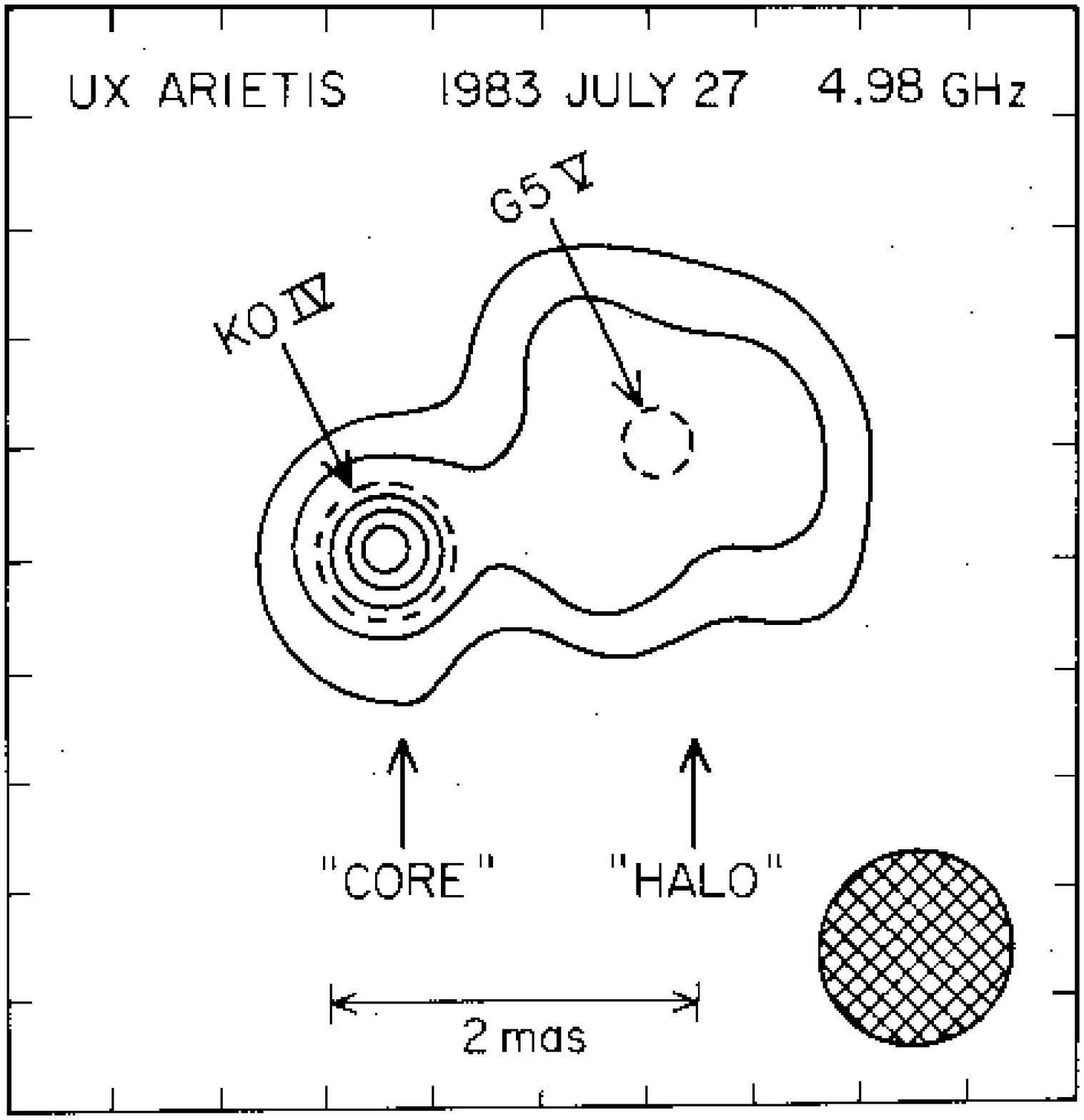}
\qquad
\includegraphics[width=5.5cm,angle=0]{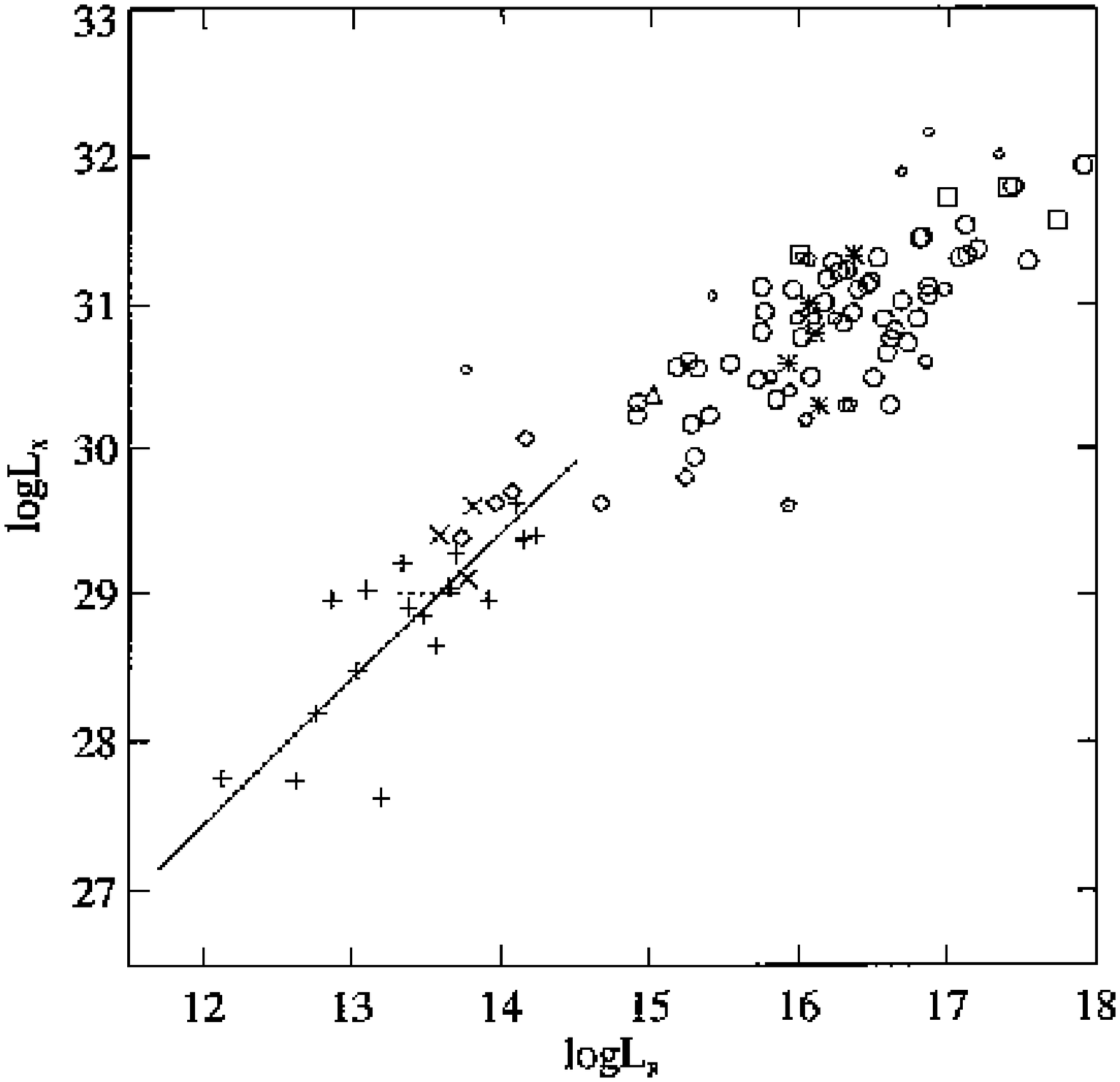}
\caption{{\it Left}: VLBI map of the RS~CVn system UX~Ari. A core-halo structure is visible (from Mutel \etal\ \cite{mutel85}). {\it Right}: Correlation between radio and X-ray luminosities of different types of active stars. The RS~CVn binaries are marked as circles (from G\"udel \& Benz \cite{gudel93}).}
\label{uxa}
\end{figure}

One of the most active sources at radio wavelengths is the system UX~Ari. The
radio emission of this system is highly variable. VLBI observations of UX Ari
have shown that also the source structure is variable. During strong flares a
compact bright source of stellar size and a fainter component of dimensions
comparable to the binary separation are observed, as can be seen in the left
image plotted in Figure~\ref{uxa}, obtained by Mutel \etal\ (1985). At lower
flux levels, however, only the extended component is present (Massi \etal\
\cite{massi88}). The $T_{\rm B}$, the sizes measured by VLBI and the moderate
circular polarization are consistent with gyrosynchrotron radiation by
electrons of a few MeV or less spiraling in fields of $\sim$10 to $\sim$100~G
(Dulk \cite{dulk85}). VLBI maps showing a clear variation of the source
structure with time have been obtained by Franciosini \etal\
(\cite{franciosini99}). Another example of the core-halo morphology is the
VLBI image of the binary system HR~5110 obtained by Ransom \etal\
(\cite{ransom03}), in which the core source has a smaller size than that of
the chromospherically active K subgiant component, whereas the halo size is
almost twice as much as the separation of the centers of the K and F stars.

A correlation between quiescent radio (5--8~GHz) and X-ray luminosity of
magnetically active stars has been found (see Figure~\ref{uxa}, right), which
suggests that radio and X-ray emissions are activity indicators that reflect
the level of magnetic activity (G\"udel \& Benz \cite{gudel93}). In
particular, for RS~CVn and Algol systems, the correlation between the radio
and the X-ray luminosities is $L_{\rm R} \propto L_{\rm X}^{1.0-1.3}$.
Assuming that the release of magnetic energy accumulated by the stretching of
magnetic lines caused by turbulent motions is a heating mechanism of coronal
loops, Chiuderi-Drago \& Franciosini (\cite{chiuderi93}) estimated that if
this energy is radiated away in the form of X-rays, then $L_{\rm X} \propto
B^{2}V$, where $B$ is the magnetic field and $V$ is the loop's volume. On the
other hand, these authors assumed a power-law electron population of index
$\delta$ in the same loop emitting optically thin gyrosynchrotron emission, so
the radio luminosity is $L_{\rm R} \propto B^{-0.22+0.96\delta}V$. From the
above relationships it follows that $L_{\rm R} \propto L_{\rm
X}^{0.45\delta-0.11}V^{1.11-0.45\delta}$. For $\delta = 3$, neglecting the
volume correction, they found $L_{\rm R} \propto L_{\rm X}^{1.24}$, in
agreement with the empirical values.

The characteristics of the radio and X-ray emission from Algol systems are
very similar to those of RS~CVn binaries. In fact, some binaries like HR~5110
show typical features of both systems, and their classification as belonging
to one class or the other is not simple. Unlike RS~CVn binaries, the radio
emission from Algol systems is usually unpolarized or very weakly polarized
even during quiescence. A VLBA image of Algol by Mutel \etal\ (\cite{mutel98})
represents the first evidence for double-lobed structure in the radio corona
of an active late-type star in the quiescent state. The individual lobes are
strongly circularly polarized and of opposite helicity, although the total
emission is only weakly polarized. The radio emission is suggested to be
gyrosynchrotron emission from optically thin emission regions containing
mildly relativistic electrons in a dipolar magnetic field. Using VLBI
astrometry, Lestrade \etal\ (\cite{lestrade93}) found that the radio emission
from Algol is related to the magnetically active K subgiant component.

\section{X-ray binaries}\label{xrb}

An X-ray binary is a binary system containing a compact object, either a
neutron star or a stellar-mass black hole, that emits X-rays as a result of a
process of accretion of matter from the companion star. Several scenarios 
have been proposed to explain this X-ray emission, depending on the nature of 
the compact object, its magnetic field in the case of a neutron star, and the 
geometry of the accretion flow. The accreted matter is accelerated to
relativistic speeds, transforming its potential energy provided by the intense
gravitational field of the compact object into kinetic energy. Assuming that
this kinetic energy is finally radiated, the accretion luminosity can be
computed, finding that this mechanism provides a very efficient source of
energy, even much higher efficiency than that for nuclear reactions.

On its way to the compact object, the accreted matter carries angular momentum
and usually forms an accretion disk around it. The matter in the disk looses
angular momentum due to viscous dissipation, which produces a heating of the
disk, and falls towards the compact object in a spiral trajectory. The black
body temperature of the last stable orbit in the case of a BH accreting at the
Eddington limit is given by:
\begin{equation}
T \sim 2 \times 10^7 M^{-1/4}
\label{eq:tlast}
\end{equation}
where $T$ is expressed in Kelvin and $M$ in $M_{\odot}$ (Rees 1984). For a
compact object of a few solar masses, $T \sim10^7$~K. At this temperature the
energy is mainly radiated in the X-ray domain.

In High Mass X-ray Binaries (HMXBs) the donor star is an O or B early type
star of mass in the range $\sim8$--$20$~$M_{\odot}$ and typical orbital
periods of several days. HMXBs are conventionally divided into two subgroups:
systems containing a B star with emission lines (Be stars), and systems
containing a supergiant (SG) O or B star. In the first case, the Be stars do
not fill their Roche lobe, and accretion onto the compact object is produced
via mass transfer through a decretion disk. Most of these systems are
transient X-ray sources during periastron passage. In the second case, OB SG
stars, the mass transfer is due to a strong stellar wind and/or to Roche lobe
overflow. The X-ray emission is persistent, and large variability is usual.
The most recent catalogue of HMXBs was compiled by Liu \etal\ (\cite{liu00}),
and contains 130 sources.

In Low Mass X-ray Binaries (LMXBs) the donor has a spectral type later than B,
and a mass $\leq2$~$M_{\odot}$. Although typically is a non-degenerated star,
there are some examples where the donor is a WD. The orbital periods are in
the range 0.2--400 hours, with typical values $<24$ hours. The orbits are
usually circular, and mass transfer is due to Roche lobe overflow. Most of
LMXBs are transients, probably as a result of an instability in the accretion
disk or a mass ejection episode from the companion. The typical ratio between
X-ray to optical luminosity is in the range $L_{\rm X}/L_{\rm
opt}\simeq100$--$1000$, and the optical light is dominated by X-ray heating of
the accretion disk and the companion star. Some LMXBs are classified as `Z'
and `Atoll' sources, according to the pattern traced out in the X-ray
color-color diagram. `Z' sources are thought to be weak magnetic field neutron
stars of the order of $10^{10}$~G with accretion rates around
0.5--1.0~$\dot{M}_{\rm Edd}$. `Atoll' sources are believed to have even weaker
magnetic fields of $\leq10^8$~G and lower accretion rates of
0.01--0.1~$\dot{M}_{\rm Edd}$. The most recent catalogue of LMXBs was 
compiled by Liu \etal\ (2001), and contains 150 sources.

Recently, Grimm \etal\ (\cite{grimm02}) estimated that the total number of
X-ray binaries in the Galaxy brighter than 2$\times 10^{34}$ erg~s$^{-1}$ is
about 705, being distributed as $\sim$325 LMXBs and $\sim$380 HMXBs.

\subsection{Radio emitting X-ray binaries (REXBs)}

The first X-ray binary known to display radio emission was Sco~X-1 in the late
1960s. Since then, many X-ray binaries have been detected at radio wavelengths
with flux densities $\geq0.1$--$1$~mJy. The flux densities detected are
produced in small angular scales, which rules out a thermal emission
mechanism. The most efficient known mechanism for production of intense radio
emission from astronomical sources is the synchrotron emission mechanism, in
which highly relativistic electrons interacting with magnetic fields produce
intense radio emission that tends to be linearly polarized. The observed radio
emission can be explained by assuming a spatial distribution of non-thermal
relativistic electrons, usually with a power-law energy distribution,
interacting with magnetic fields.

Since some REXBs, like SS~433, were found to display elongated or jet-like
features, like in AGN and quasars, it was proposed that flows of relativistic
electrons were ejected perpendicular to the accretion disk, and were
responsible for synchrotron radio emission in the presence of a magnetic
field. Models of adiabatically expanding synchrotron radiation-emitting
conical jets may explain some of the characteristics of radio emission from
X-ray binaries (Hjellming \& Johnston \cite{hjellming88}). Several models have
been proposed for the formation and collimation of the jets, including the
presence of an accretion disk close to the compact object, a magnetic field in
the accretion disk, or a high spin for the compact object. However, there is
no clear agreement on what mechanism is exactly at work.

There are eight radio emitting HMXBs and 35 radio emitting LMXBs. Since the
strong magnetic field of the X-ray pulsars disrupts the accretion disk at
several thousand kilometers from the neutron star, there is no inner accretion
disk to launch a jet and no synchrotron radio emission has ever been detected
in any of these sources. Although the division of X-ray binaries in HMXBs and
LMXBs is useful for the study of binary evolution, it is probably not
important for the study of the radio emission in these systems, where the only
important aspect seems to be the presence of an inner accretion disk capable
of producing radio jets. However, the eight radio emitting HMXBs include six
persistent and two transient sources, while among the 35 radio emitting LMXBs
we find 11 persistent and 24 transient sources. The difference between the
persistent and transient behavior clearly depends on the mass of the donor.

Excluding X-ray pulsars, $20$ to $25\%$ of the catalogued galactic X-ray
binaries have been detected at radio wavelengths regardless of the nature of
the donor. The corresponding ratio of detected/observed sources is probably
much higher. However, it is difficult to give reliable numbers, since
observational constrains arise when considering transient sources observed in
the past (large X-ray error boxes, single dish and/or poor sensitivity radio
observations, etc.), and likely many non-detections have not been published.

\subsection{Microquasars}

\begin{figure}
\center 
\includegraphics[width=6cm]{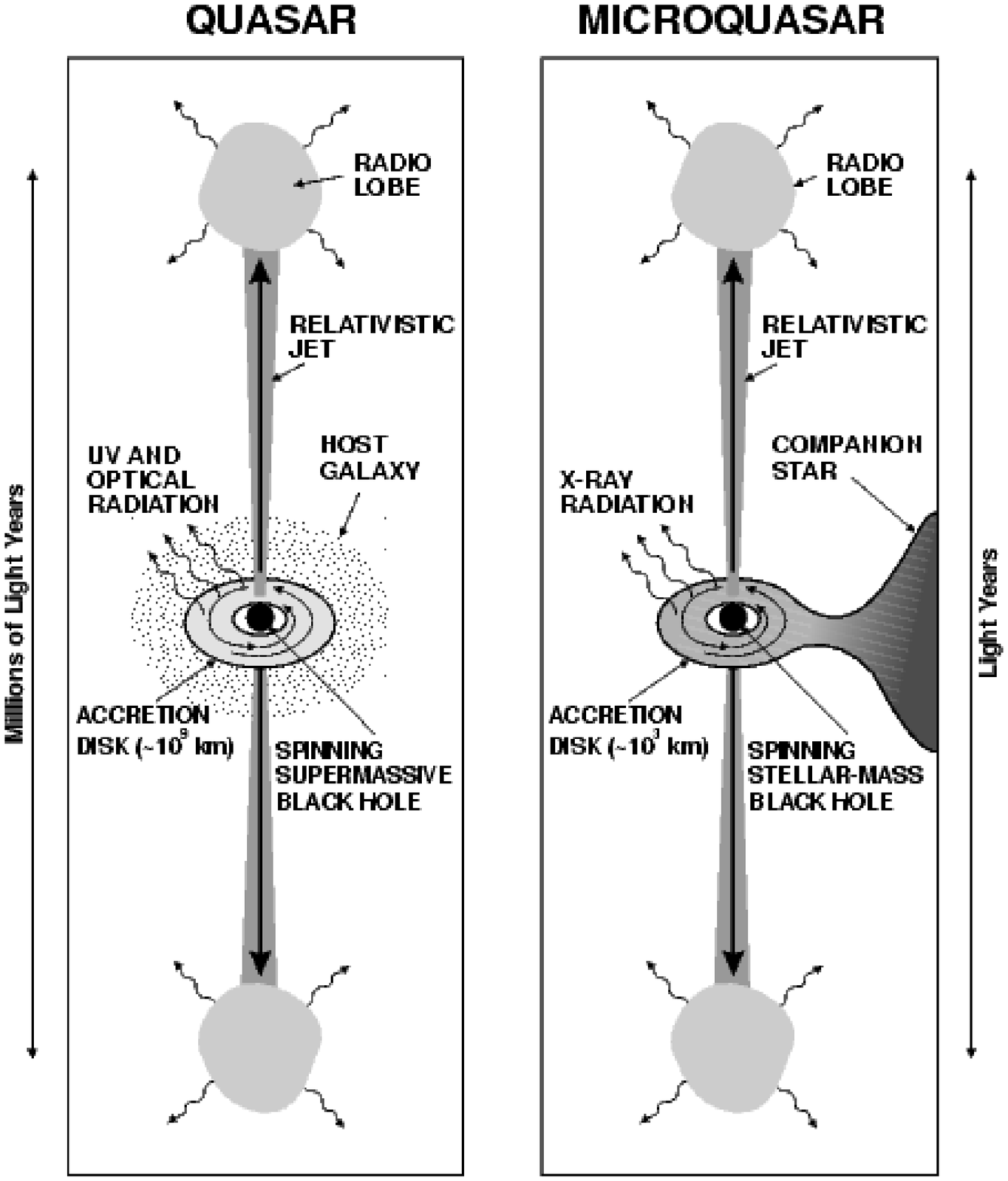}
\caption{Comparative illustration of the analogy between quasars and microquasars. Note the extreme differences in the order of magnitude of the physical parameters involved (from Mirabel \& Rodr\'{\i}guez \cite{mirabel98}).}
\label{qmq}
\end{figure}

A microquasar is a radio emitting X-ray binary displaying relativistic radio
jets. The name was given not only because of the observed morphological
similarities between these sources and the distant quasars but also because of
physical similarities, since when the compact object is a black hole, some
parameters scale with the mass of the central object (Mirabel \&
Rodr\'{\i}guez \cite{mirabel99}). A schematic illustration comparing some
parameters in quasars and microquasars is shown in Figure~\ref{qmq}.

From Eq.~\ref{eq:tlast}, a typical temperature of a microquasar containing a
stellar-mass black hole is $T\sim10^7$~K, while that of a quasar containing a
supermassive black hole ($10^7$--$10^9$~$M_{\odot}$) is $T\sim10^5$~K. This
explains why in microquasars the accretion luminosity is radiated in X-rays, 
while in quasars it is radiated in the optical/UV domain. The characteristic
jet sizes seem to be proportional to the mass of the black hole. Radio jets in
microquasars have typical sizes of a few light years, while in quasars may
reach distances of up to several million light years. The timescales are also
directly scaled with the mass of the black hole following $\tau\simeq R_{\rm
S}/c=2GM_{\rm X}/c^3\propto M_{\rm X}$. Therefore, phenomena that take place
in timescales of years in quasars can be studied in minutes in microquasars.
Thus, microquasars mimic, on smaller scales, many of the phenomena seen in
AGNs and quasars, but allow a better and faster progress in the understanding
of the accretion/ejection processes that take place near compact objects. 

The current number of microquasars is $\sim$16 among the 43 catalogued REXBs.
Some authors (Fender 2001) have proposed that all REXBs are microquasars, and
would be detected as such provided that there is enough sensitivity and/or
resolution in the radio observations. The known microquasars, compiled from
different sources, are listed in Table~\ref{census}.

\begin{table}
{\small
\caption[]{\label{census} {\bf Microquasars in our Galaxy}}
\begin{tabular}{@{}l@{\hspace{0.07cm}}c@{\hspace{0.05cm}}cc@{\hspace{0.05cm}}c@{\hspace{0.05cm}}c@{\hspace{0.05cm}}c@{\hspace{0.05cm}}c@{\hspace{0.05cm}}c@{}}
             &                 &           &      &              &              &                           \\ 
\hline \noalign{\smallskip}
Name $\&$ Position &  System  & $D$ & $P_{\rm orb}$  & $M_{\rm compact}$   & Activity & $\beta_{\rm apar}$ & $\theta$$^{\rm (c)}$ & Jet size  \\
~~~(J2000.0)  & type$^{\rm (a)}$& (kpc) & (d)   &   $(\mo)$  & radio$^{\rm (b)}$&   &  & (AU)\\ 
\noalign{\smallskip} \hline \noalign{\smallskip}
\multicolumn{9}{c}{\bf High Mass X-ray Binaries (HMXB)}\\
\noalign{\smallskip} \hline \noalign{\smallskip}

{\bf LS~I~+61~303} & B0V & 2.0 & 26.5 & $-$ & p & $\geq$0.4 & $-$ & 10$-$700 \\
$02^{\rm h}40^{\rm m}$31 \rl 66 & +NS? &  & & & & & \\
$+61^{\circ}13^{\prime}$45\pri 6 & & & & & & & & \\

{\bf V4641~Sgr}  & B9III & $\sim10$ & 2.8  & 9.6 & t & $\ge9.5$ & $-$&$-$ \\
$18^{\rm h}19^{\rm m}$21\rl 48 &+BH & \\
$-25^{\circ}25^{\prime}$36\pri 0 &     &       &   &  &  &         \\ 
 
{\bf LS~5039}    &  O6.5V((f)) & 2.9 & 4.4 & 1$-$3 & p  & $\geq0.15$ &$<81^{\circ}$& 10$-$1000 \\
$18^{\rm h}26^{\rm m}$15\rl 05 & +NS?& &   & & & & &  \\
$-14^{\circ}50^{\prime}$54\pri 24 &   &       &  &  &  \\ 
  
{\bf SS~433} & evolved A?   & 4.8  &  13.1 &  11$\pm$5?& p & 0.26 &  $79^{\circ}$&$\sim10^4$$-$$10^6$    \\
$19^{\rm h}11^{\rm m}$49\rl 6 &+NS  & &   & & & & & \\
$+04^{\circ}58^{\prime}58^{\prime\prime}$ &   &    &  &  &   &  &  &    \\
  
{\bf Cygnus~X-1} & O9.7Iab  & 2.5 & 5.6 & 10.1 &  p &$-$& 40$^{\circ}$& $\sim40$\\
$19^{\rm h}58^{\rm m}$21\rl 68 & +BH & &   & & & & & \\
$+35^{\circ}12^{\prime}$05\pri 8 &   &    &  &  &   &  &  &      \\
  
{\bf Cygnus~X-3} &  WNe     &  9      &  0.2   & $-$& p  & 0.69 & 73$^{\circ}$ & $\sim10^4$    \\
$20^{\rm h}32^{\rm m}$25\rl 78 & +BH? & &   & & & & & \\
$+40^{\circ}57^{\prime}$28\pri 0     &       &    &    &      &   &  & & \\

\noalign{\smallskip} \hline \noalign{\smallskip}
\multicolumn{9}{c}{\bf Low Mass X-ray Binaries (LMXB)}\\
\noalign{\smallskip} \hline \noalign{\smallskip}

{\bf XTE~J1118+480}   & K7V$-$M0V    & 1.9  &  0.17  & 6.9$\pm$0.9 &  t &$-$& $-$&$\le0.03$   \\
$11^{\rm h}18^{\rm m}$10\rl 85 & +BH \\
$+48^{\circ}02^{\prime}$12\pri 9          &       &    &    &   \\ 
      
{\bf Circinus~X-1}    &  Subgiant &  5.5     &  16.6  &$-$&  t   & $\geq0.1$& $>70^{\circ}$ & $>10^4$ \\
$15^{\rm h}20^{\rm m}$40\rl 9 & +NS \\
$-57^{\circ}10^{\prime}01^{\prime\prime}$       \\ 
 
{\bf XTE~J1550$-$564} &   G8$-$K5V & 5.3  & 1.5   & 9.4 &  t &$>2$& $-$& $\sim10^3$     \\
$15^{\rm h}50^{\rm m}$58\rl 70 & +BH & &   & & & & &  \\
$-56^{\circ}28^{\prime}$35\pri 2   &       &      &      & &    \\ 
 
{\bf Scorpius~X-1}     & Subgiant    & 2.8      &  0.8  &  1.4 &  p &$ 0.68$& $44^{\circ}$& $\sim40$    \\
$16^{\rm h}19^{\rm m}$55\rl 1 & +NS  \\
$-15^{\circ}38^{\prime}25^{\prime\prime}$ &  &      &     &    &   &    \\
  
{\bf GRO~J1655$-$40} & F5IV  & 3.2    &  2.6   & 7.02 & t & 1.1& $72^{\circ}$$-$$85^{\circ}$&  8000     \\
$16^{\rm h}54^{\rm m}$00\rl 25  & +BH & &   & & & & &  \\
$-39^{\circ}50^{\prime}$45\pri 0 &   &      &        &     &     & &    \\ 
   
{\bf GX~339$-$4}   &   $-$         & $\sim4$     &  1.76  & 5.8$\pm$0.5 & t& $-$&$-$& $<$ 4000    \\
$17^{\rm h}02^{\rm m}$49\rl 5 & +BH \\
$-48^{\circ}47^{\prime}23^{\prime\prime}$ &   &       &        &  &    &   \\  
  
{\bf 1E~1740.7$-$2942}& $-$  & 8.5? &  12.5?   &$-$ & p &$-$& $-$&$\sim10^6$  \\
$17^{\rm h} 43^{\rm m} 54$\rl 83 & +BH ?& &   & & & & & \\
$-29^{\circ} 44^{\prime}$42\pri 60 &   &     &     &   &      \\ 
 
{\bf XTE~J1748$-$288} &  $-$ &  $\geq8$   &  ?    &   $>4.5$? & t & 1.3&$-$ & $>10^4$         \\
$17^{\rm h}48^{\rm m}$05\rl 06  &+BH? \\
$-28^{\circ}28^{\prime}$25\pri 8 &         &    &   & &   \\ 
    
{\bf GRS~1758$-$258}  &  $-$ & 8.5?  & 18.5?  &$-$   & p &$-$& $-$&$\sim10^6$\\
$18^{\rm h}01^{\rm m}$12\rl 40  &+BH ?\\
$-25^{\circ}44^{\prime}$36\pri 1 &    &      &     &    &      \\
   
{\bf GRS~1915+105}    &  K$-$M III   & 12.5 & 33.5  & 14$\pm$4 &  t & 1.2$-$1.7& $66^{\circ}$$-$$70^{\circ}$&$\sim10$$-$$10^4$ \\
$19^{\rm h}15^{\rm m}$11\rl 55 &+BH & &   & & & & & \\
$+10^{\circ}56^{\prime}$44\pri 7 &     &       &    &&& & &       \\
\hline
\end{tabular}
{\small
Notes: $^{\rm (a)}$ NS: neutron star; BH: black hole. $^{\rm (b)}$ p: persistent; t:transient. $^{\rm (c)}$ jet inclination.
}
}
\end{table}

\subsubsection{Relativistic effects}

The modern interferometers, working at radio wavelengths, are the only
instruments that have provided a direct view of the most spectacular phenomena
in microquasars. Their angular resolution well below the sub-arcsecond level
allow us to follow the path and brightness decay of plasma clouds (plasmons)
ejected along the relativistic jets into opposite directions. Due to the huge
velocities involved, the effects predicted by the theory of Special Relativity
must be taken into account for a correct interpretation of the observed data.
Among them, we have the illusion of superluminal motion, as well as the
difference in apparent brightness between the approaching and the receding jet
due to relativistic aberration of light. \\

a) {\it Apparent superluminal motion:} \\

\begin{figure} 
\center
\includegraphics[width=5cm]{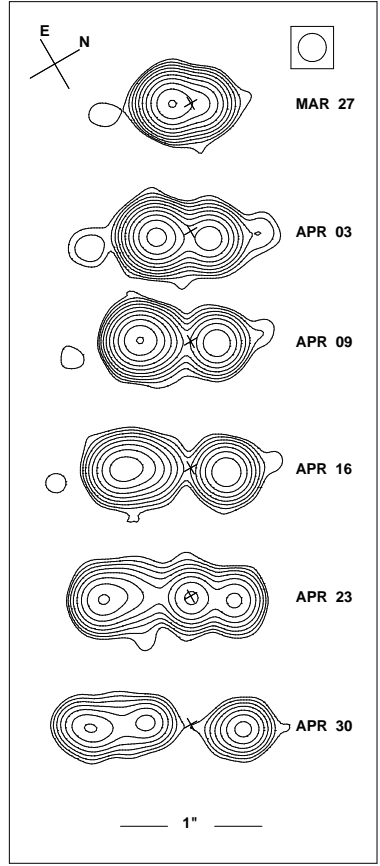}
\caption{VLA multiepoch observations at 3.6~cm wavelength of the superluminal 
ejection of GRS~1915+105 in 1994. The approaching component, to the left, appeared to move at $1.25c$ and be brighter than the receding one, to the right, which moved with an apparent velocity of $0.65c$ (from Mirabel \& Rodr\'{\i}guez \cite{mirabel94}).} 
\label{suplum}
\end{figure}

Let us assume that a source ejects two identical plasma clouds into opposite
directions. The illusion of superluminal motion occurs for an ejection
velocity $v=\beta c$ close to the speed of light and a rather small angle
$\theta$ with respect to the line of sight. The fact that the approaching
condensation reduces its distance to the observer by an amount $v t
\cos{\theta}$ makes the light travel time towards the observer progressively
shorter. The apparent velocity measured is thus given by:
\begin{equation}
v_{\rm a,r} = { v \sin{\theta} \over  ( 1 \mp \beta \cos{\theta} ) },
\end{equation}
for the approaching and the receding clouds, respectively. The minus sign, 
corresponding to the approaching case, may lead to a value of $v_{\rm a}$
arbitrarily large provided that $\beta$ and $\cos{\theta}$ approach unity. 
The first galactic object observed with superluminal motion was the
microquasar GRS~1915+105 (Mirabel \& Rodr\'{\i}guez \cite{mirabel94}). Until
then, such relativistic illusion had been observed only in extragalactic
objects like quasars. Figure~\ref{suplum} shows the time evolution of a
GRS~1915+105 eruption in 1994, a bipolar ejection of two plasmons going away
from the central core. Assuming a kinematical distance estimate to
GRS~1915+105 of 12.5~kpc, the observed proper motions of the plasma clouds
translate into apparent velocities of 1.25 and 0.65 times the speed of light.
Other confirmed cases of superluminal sources in the Galaxy are the
microquasars GRO~J1655$-$40, XTE~J1748$-$288, V4641~Sgr and XTE~J1550$-$564.
\\

b) {\it Relativistic aberration:} \\

This effect is usually know as Doppler boosting. As above, let us assume that 
the two plasma clouds ejected into opposite directions are identical, with a
radiation flux density $S_0$ in their respective reference frame. The
synchrotron spectrum of each cloud, as a function of frequency $\nu$, follows
a power-law $S_0 \propto \nu^{\alpha}$, being $\alpha$ the so-called spectral
index (usually $\alpha \simeq -0.7$). In the reference frame of the observer,
the resulting flux density appears to be different from $S_0$. Let $S_{\rm
a,r}$ be the flux density observed from the approaching and the receding
clouds, respectively. The relationship between the emitted and observed flux
density is given by:
\begin{equation}
S_{\rm a,r} = { S_0 \over \left[ \Gamma ( 1 \mp \beta \cos{\theta} ) \right]^{k-\alpha} },
\end{equation}
where $\Gamma = 1/\sqrt{1-\beta^2}$ is the bulk Lorentz factor, and the
constant $k$ takes the values 3 or 2 depending on the case of discrete clouds
or a continuous jet, respectively.

If $\theta$ is small ($\leq 10^{\circ}$) and $\beta$ close to unity, the
brightness of the approaching cloud is considerably boosted and it may look
thousands of times brighter than the receding one. This is the so-called
Doppler favouritism, which only allows to see one side of the jet
condensations in very distant quasars, where a strong amplification is needed
for the emission to be detectable. This effect is shown in
Figure~\ref{suplum}, where the approaching jet seems to be faster and clearly
brighter than the opposite counter-jet.

\subsubsection{Accretion disk and jet ejection}

The theoretical models attempting to understand the jet formation and its
connection with the accretion disk had a seminal contribution in the works by
Blandford \& Payne (\cite{blandford82}). These authors explored the
possibility of extracting energy and angular momentum from the accretion disk
by means of a magnetic field whose lines extend towards large distances from
the disk surface. Their main result was the confirmation of the theoretical
possibility to generate a flow of matter from the disk itself towards outside,
provided that the angle between the disk and the lines was smaller than
60$^{\circ}$. Later on, the flow of matter is collimated at large distances
from the disk by the action of a toroidal component of the magnetic field. In
this way, two opposite jets could be formed flowing away perpendicularly to
the accretion disk plane.

\begin{figure} 
\center
\includegraphics[width=6cm]{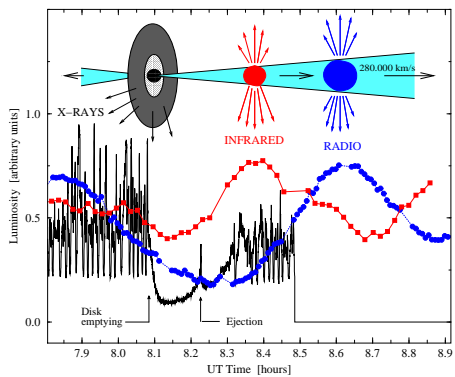}
\caption{Multi-wavelength behaviour of the microquasar GRS~1915+105 as observed in September 8th 1997 (Mirabel \etal\ \cite{mirabel98}). The radio data at 3.6~cm (grey squares) were obtained with the VLA interferometer; the infrared observations at 2.2 micron (black squares) are from the UKIRT; the continuous line is the X-ray emission as observed by RXTE in the 2--50 keV range.} 
\label{grs1915}
\end{figure}

To confirm observationally the link between accretion disk and the genesis of
the jets is by no means an easy task. The collimated ejections in GRS~1915+105
provide one of the best studied cases supporting the proposed disk/jet
connection. In Figure~\ref{grs1915}, from Mirabel \etal\ (\cite{mirabel98}),
simultaneous observations are presented at radio, infrared and X-ray
wavelengths. The data show the development of a radio outburst, with a peak
flux density of about 50~mJy, as a result of a bipolar ejection of plasma
clouds. However, previous to the radio outburst there was a clear precursor
outburst in the infrared. The simplest interpretation is that both flaring
episodes, radio and infrared, were due to synchrotron radiation generated by
the same relativistic electrons of the ejected plasma. The adiabatic expansion
of plasma clouds in the jets causes losses of energy of these electrons and,
as a result, the spectral maximum of their synchrotron radiation is
progressively shifted from the infrared to the radio domain.

It is also important to note the behaviour of the X-ray emission in
Figure~\ref{grs1915}. The emergence of jet plasma clouds, that produces the
infrared and radio flares, seems to be accompanied by a sharp decay and
hardening of the X-ray emission (8.08--8.23~h UT in the figure). The X-ray
fading is interpreted as the disappearance, or emptying, of the inner regions
of the accretion disk (Belloni \etal\ \cite{belloni97}). Part of the matter
content in the disk is then ejected into the jets, perpendicularly to the
disk, while the rest is finally captured by the central black hole.
Additionally, Mirabel \etal\ (\cite{mirabel98}) suggest that the initial time
of the ejection coincides with the isolated X-ray spike just when the hardness
index suddenly declines (8.23~h UT). The recovery of the X-ray emission level
at this point is interpreted as the progressive refilling of the inner
accretion disk with a new supply of matter until reaching the last stable
orbit around the black hole.

This behaviour in the light curves of GRS~1915+105 has been repeatedly
observed by different authors (e.g. Fender \etal\ \cite{fender97}; Eikenberry
\etal\ \cite{eikenberry98}), providing thus a solid proof of the so-called
disk/jet symbiosis in accretion disks. All the observed events took less than
half an hour to occur, and their equivalent in quasars, or AGNs, would require
a much longer time span, a minimum of some few years. Despite the complexity
in the GRS~1915+105 light curves, the episodes of X-ray emission decay with
associated hardening are reminiscent of the well known low/hard state typical
of persistent black hole candidates (Cygnus~X-1, 1E~1740.7$-$2942,
GRS~1758$-$258 and GX~339$-$4). The transitions towards this state are often
accompanied by radio emission with flat spectrum, interpreted as due to the
continuous creation of compact synchrotron jets partially self-absorbed.

It is worth mentioning the observational work by Marscher \etal\
(\cite{marscher02}), who presented evidence of the disk/jet symbiosis also in
an AGN, the active galaxy 3C~120. Using VLBI techniques they observed episodes
of ejection of superluminal plasma just after the decay and hardening of the
X-ray emission. This is precisely the same behaviour displayed by
GRS~1915+105. The events in 3C~120 seem to be recurrent with an interval of
one year, which is consistent with a mass of the compact object of
$\sim10^{7}$ $M_{\odot}$. Such observations strongly support the idea of
continuity between galactic microquasars and AGNs in the Universe.

\subsubsection{Precession}

\begin{figure}
\center 
\includegraphics[width=5.5cm]{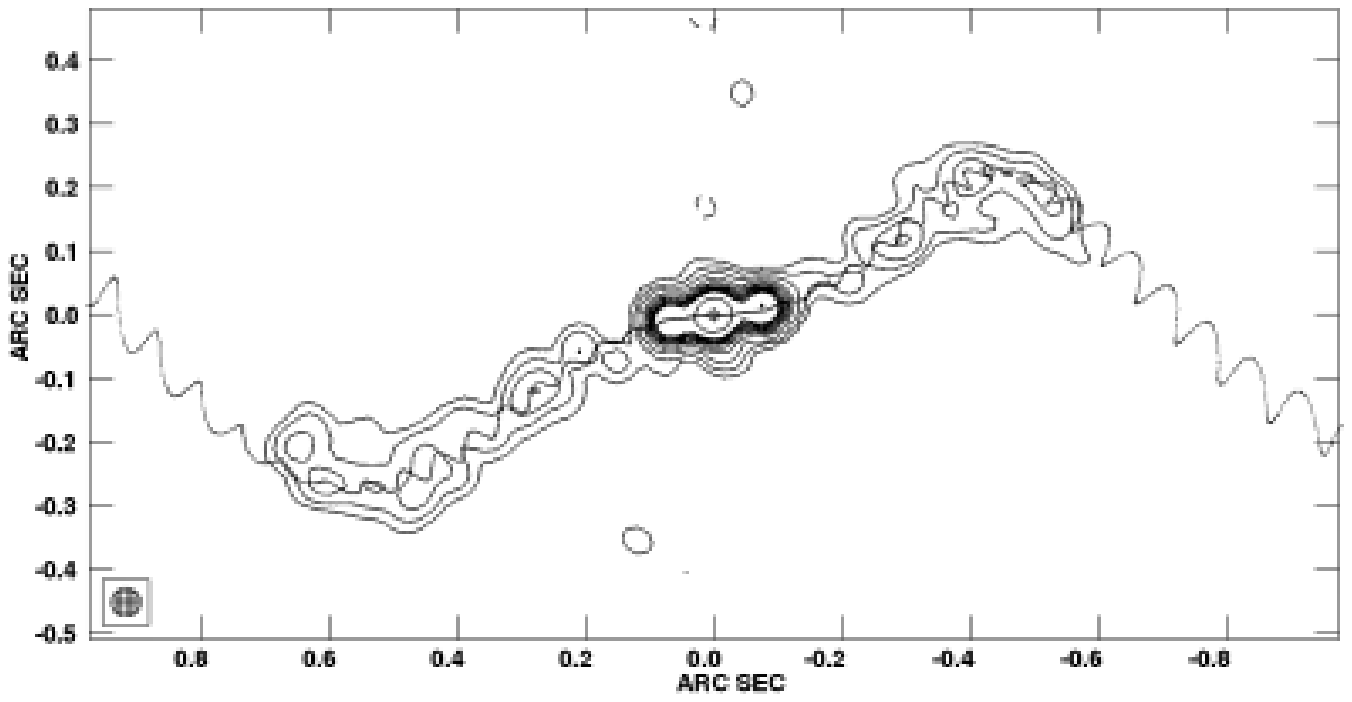}
\qquad
\includegraphics[width=5.5cm]{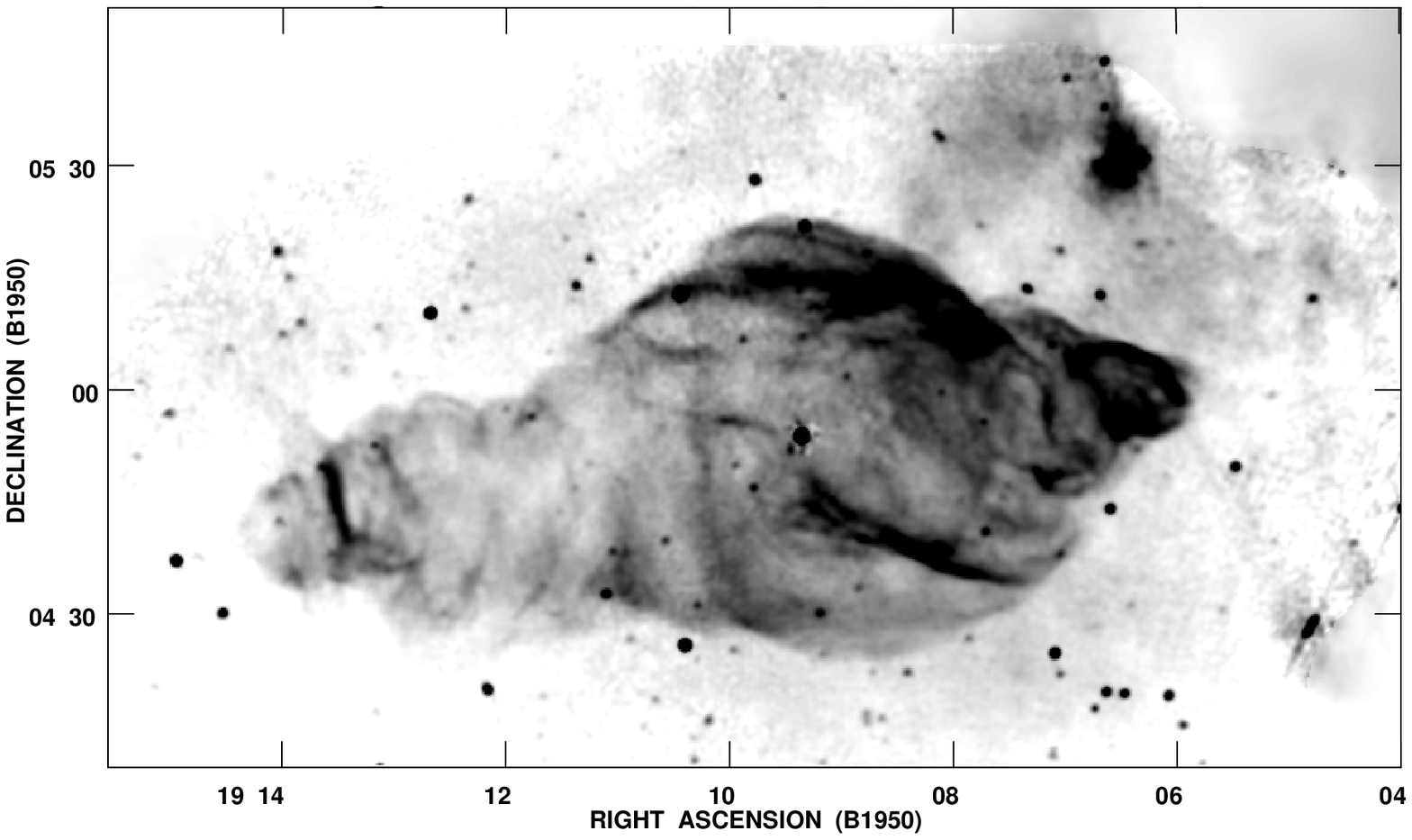}
\caption{{\it Left}: 5~GHz image of SS~433 obtained with MERLIN. A kinematic model locus has been plotted (from Stirling \etal\ \cite{stirling02}). 
{\it Right}: VLA image at 1465~MHz of the W50 nebula surrounding SS~433 at the center. Traces of the interaction of the jets of SS~433 with the surrounding gas are visible (from Dubner \etal\ \cite{dubner98}).}
\label{ss433} 
\end{figure}

Historically, the first microquasar discovered was SS~433 (see Margon 1984). 
For many years it was considered a mere curiosity in the galactic fauna. Its
plasma jets are ejected into the interstellar space at a speed of 0.26c, and
precess with a period of 163 days. The flight of plasma clouds along the jets
can be followed spectroscopically by means of their emission lines, whose
redshift or blueshift agrees with a simple model of conical precession. SS~433
is the only microquasar where such lines have been detected so far, thus
demonstrating the barionic nature of the ejecta at least in one case. The
kinematic twin-jet model for SS~433 explains not only the radial velocity
behaviour of the optical jets, but also the details of the proper motions of
radio jets to scales of a few arc seconds (Figure~\ref{ss433}) (Hjellming \&
Johnston \cite{hjellming81}; Stirling \etal\ \cite{stirling02}). Other
microquasars, such as LS~I~+61~303, V4641~Sgr, GRO~J1655$-$40 and
GRS~1915+105, might also be precessing systems, although this possibility
needs further confirmation.

\subsubsection{Strong radio events}

Cygnus~X-3 has been the subject of intensive study during the last decades, 
specially after the discovery of its giant radio outbursts in 1972 (Gregory
\etal\ \cite{gregory72}). During strong outbursts, its radio emission rises up
to three orders of magnitude in just a few days. As a result, radio jets
moving with relativistic speeds are formed. Several authors have provided 
strong observational evidence indicating that flaring radio emission
originates in expanding collimated jet-like structures (see, e.g., Mart\'{\i}
\etal\ \cite{marti01}, and references therein). The highest resolution maps of
the ejecta have been provided by Mioduszewski \etal\ (\cite{mioduszewski01}),
who observed Cygnus~X-3 with VLBA soon after a giant outburst event in 1997.
Further support for the jet scenario comes from the agreement between the
radio light curves and the predictions from theoretical models of synchrotron
emitting radio jets (see e.g. Hjellming \& Johnston \cite{hjellming88};
Mart\'{\i} \etal\ \cite{marti92}).

\subsubsection{High energy emission}

The instrument COMPTEL on board the Compton Gamma-ray Observatory detected
some microquasars at energies of several MeV. For example, Cygnus~X-1 was
detected several times and it may be even brighter above 1~MeV in the
soft/high state (McConnell \etal\ \cite{mcconnell00}). GRO~J1655$-$40 was also
detected up to $\sim$1~MeV (Grove \etal\ 1998). In the extreme energy range of
TeV $\gamma$-rays, a flux of the order of 0.25~Crab was detected from
GRS~1915+105 during the period May-July 1996, when the source was in an active
state (Aharonian \& Heinzelmann \cite{aharonian98}). However, this result
needs further confirmation given the marginal confidence of the detection.

Microquasars appear as a possible explanation for some of the unidentified
sources of high energy $\gamma$-rays detected by the experiment EGRET on board
the satellite COMPTON-GRO. The possible association between the microquasar
LS~5039 and the high energy ($E>$100~MeV) $\gamma$-ray source 3EG~J1824$-$1514
provides observational evidence that microquasars could also be sources of
high energy $\gamma$-rays (Paredes \etal\ \cite{paredes00}). LS~I~+61~303 has
also been proposed to be associated with the $\gamma$-ray source 2CG~135+01
(=3EG~J0241+6103) (Kniffen \etal\ \cite{kniffen97}).

\section{Summary and prospects}

The study of radio emission from the stars has advanced significantly over 
the last decades, both theoretically and observationally, thanks to the
development of large and sensitive interferometers. We are now able to measure
the mass loss rates in hot stars, to estimate the magnetic fields through
polarization measurements, to map stellar structures at milliarcsecond level,
and to study the processes involved in the particle energization. The
mechanisms for producing the stellar radio emission are now better understood.
Different energy sources have been invoked to explain the stellar radio
emission. For RS~CVn systems the energy source is the magnetic field, whereas
for X-ray binaries and microquasars it is the mass accretion onto the compact
companion. The study of radio stars also contributes to issues not directly
related to the astrophysical phenomena that characterize these sources, as the
matching between the radio and the optical reference systems.

Some aspects of the interpretation of the radio emission in active close
binaries remain still unclear or need additional observational support. 
Further advances in the understanding of stellar radio emission are limited 
by sensitivity and resolution. Furthermore, only a few hundred stellar radio 
sources are currently known. In this sense, the development of the Square
Kilometer Array (SKA) and the Expanded Very Large Array (EVLA) in the near
future, interferometers with much better sensitivity and spatial resolution as
well as larger frequency coverage than current ones, will undoubtedly prompt a
tremendous advance in stellar radio astronomy. For instance, the number of
stellar radio sources is expected to increase at least by about four orders of
magnitude, providing thus a much larger census of properties that will allow
us to better address fundamental questions in astronomy. Also, the discovery
of new classes of radio stars will likely bring unexpected phenomena to our
attention.

\acknowledgements
J.~M.~P. acknowledge partial support by DGI of the Ministerio de Ciencia y Tecnolog\'{\i}a (Spain) under grant AYA2001-3092, as well as partial support by the European Regional Development Fund (ERDF/FEDER). I am indebted to Joan Garc\'{\i}a-S\'anchez and Marc Rib\'o for a careful reading of the manuscript and their valuable comments.


\end{document}